\documentclass[12pt]{iopart}
\usepackage{amssymb}
\usepackage{bm}
\usepackage[numbers]{natbib}
\usepackage{graphicx}
\usepackage{booktabs}
\usepackage{siunitx}
\usepackage{subcaption}
\usepackage{hyperref}
\usepackage{xcolor}
\usepackage{soul}

\newcommand{\argmin}{\mathop{\mathrm{argmin}}}

\begin{document}

\title{GroupRegNet: A Groupwise One-shot Deep Learning-based 4D Image Registration Method}

\author{\textsuperscript{1}Yunlu Zhang, \textsuperscript{1}Xue Wu, \textsuperscript{1,2,3}H. Michael Gach, \textsuperscript{1,2}Harold Li, \textsuperscript{1,2}Deshan Yang}

\address{Departments of \textsuperscript{1}Radiation Oncology, \textsuperscript{2}Biomedical Engineering, and \textsuperscript{3}Radiology, Washington University in Saint Louis, St. Louis, MO, USA, 63110}
\ead{yangdeshan@wustl.edu}
\vspace{10pt}
\begin{indented}
\item[]September 2020
\end{indented}

\begin{abstract}
Accurate deformable 4-dimensional (4D) (3-dimensional in space and time) medical images registration is essential in a variety of medical applications. 
Deep learning-based methods have recently gained popularity in this area for the significantly lower inference time. However, they suffer from drawbacks of non-optimal accuracy and the requirement of a large amount of training data. A new method named GroupRegNet is proposed to address both limitations. The deformation fields to warp all images in the group into a common template is obtained through one-shot learning. The use of the implicit template reduces  bias and accumulated error associated with the specified reference image. The one-shot learning strategy is similar to the conventional iterative optimization method but the motion model and parameters are replaced with a convolutional neural network (CNN) and the weights of the network.
GroupRegNet also features a simpler network design and a more straightforward registration process, which eliminates the need to break up the input image into patches. The proposed method was quantitatively evaluated on two public respiratory-binned 4D-CT datasets. The results suggest that GroupRegNet outperforms the latest published deep learning-based methods and is comparable to the top conventional method pTVreg. To facilitate future research, the source code is available at \url{https://github.com/vincentme/GroupRegNet}. 
\end{abstract}

\submitto{Physics in Medicine \& Biology}

\section{Introduction}
4-dimensional (4D) (3-dimensional in space and time) medical images have been used in a variety of medical applications. For instance, 4D computed tomography (4D-CT) images have been used to determine patient-specific tumor motion patterns through deformable image registration (DIR), which is a critical step in the planning and delivery of radiation therapy of lung cancer. The subsequent tumor response and anatomy change during treatment can then be studied by registering newly acquired longitudinal volumetric scans. 

Numerous research studies have been devoted to developing accurate DIR algorithms. However, as limited by image noise, the lack of features, the use of multiple imaging modalities, and often irregular patient respiratory motion patterns, a generalized, accurate, robust, and computational efficient DIR algorithm has yet to be developed. 

DIR algorithms can be roughly categorized into two types: conventional and learning-based methods. The conventional methods \cite{fu2018adaptive,papiez2018fast,wu2013estimating} formulate the registration problem as an iterative optimization problem while the learning-based methods generate a regression model from the training data. 
Recently, deep learning-based registration methods \cite{fu2020lungregnet,fechter2020one,jiang2020multi,sentker2018gdl} have gained popularity due to their low inference time. Like many subareas in computer vision, the features learned through training have shown to be more robust and more general than handcrafted features. 
In brief, the learning-based methods can be classified as supervised or unsupervised. The former requires a large amount of annotated segmentation data or artificial deformation data to train the network. However, generating such data is time-consuming and thus often impractical, limited to a specific problem, and error-prone. Therefore, most of the recently proposed learning-based methods adopted the unsupervised approach that is guided by the similarity loss. However, these unsupervised learning methods have not yet achieved the accuracy of a few conventional methods \cite{fu2020lungregnet}.

In addition to accuracy, most deep learning-based methods require a large amount of high-quality training data. Even for unsupervised methods, a sufficient number of training images of the same modality is required for training; this is however often not possible for many medical applications. The recent developed one-shot learning strategy \cite{fechter2020one} eliminated this constraint while achieving excellent results. The one-shot learning strategy is similar to classical registration methods but replaces the conventional motion model and its parameters with a convolutional neural network (CNN) and its weights where the weights are trained from scratch only using the images to be registered.

Another strategy that can be beneficial is via groupwise registration that registers multiple images to a common space instead of in pairs, and is especially suitable for 4D-CT registration. Three variations of groupwise registration exist in the literature: reference-based, sum-of-pairs \cite{papiez2018fast}, and implicit template approach \cite{wu2013estimating}. In specific, the reference-based approach requires the selection of one particular image as reference, the sum-of-pairs approach attempts to reduce the losses among all pairs of images, and the implicit template approach is able to avoid the bias caused by selecting one particular image as reference while being computationally efficient comparing to the sum-of-pairs approach. 

In this study, an unsupervised deep learning-based DIR method that employs both groupwise registration and one-shot strategy, GroupRegNet, is proposed to register 4D medical images and then to determine all pairwise deformation vector fields (DVFs). The major contributions of this work are summarized as follows. First, groupwise registration with implicitly determined template image strategy is implemented using a neural network. The periodic motion in the 4D image group is also utilized through cyclic loss. Second, the one-shot unsupervised learning approach eliminates the need for abundant training data. Third, the proposed method features a simpler network design, a minimal prepossessing, and a straightforward registration process compared to other learning-based DIR methods. In terms of accuracy, the GroupRegNet method outperforms the latest published deep learning-based methods and is comparable to a top conventional method pTVreg. 

\section{Methods}

\subsection{Problem formulation}
Let $I^N$ denotes a group of gray scale images $I^N=\{I_n|n=1,\ldots,N\}$. $I_n:\Omega\rightarrow\mathbb{R},\Omega\subset\mathbb{R}^d$ represents each image in the group. The proposed method applies for $I_n$ as 2D or 3D images, but throughout the rest of the paper, we assume they are 3D images representing one phase in time in a 4D-CT dataset. The objective of GroupRegNet is to find a set of dense transformations that map the same anatomical locations between any two individual images in the group. 

The optimization problem to be solved by GroupRegNet is formulated as: 
\begin{equation}\label{eq:1}
\argmin_{T^N_{\mathrm{tem}}} (L_{\mathrm{simi}}(T^N_{\mathrm{tem}}\circ I^N,I_{tem})+\lambda_0 L_{\mathrm{smo}}(T^N_{\mathrm{tem}}))+\lambda_1 L_{\mathrm{cyc}}(T^N_{\mathrm{tem}})),
\end{equation}
where $L_{\mathrm{simi}}$, $L_{\mathrm{smo}}$, and $L_{\mathrm{cyc}}$ are the similarity, smoothness, and cyclic regularization losses, $T^N_{\mathrm{tem}}$ is a set of transformations $\{T^n_{\mathrm{tem}}|n=1,\ldots,N\}$ that maps anatomical locations in the template to the corresponding locations in the input images, $T^n_{\mathrm{tem}}\circ I_n$ and $T^N_{\mathrm{tem}}\circ I^N$ represent the warped $n$th input image and all warped input images, respectively, $I_{tem}=\frac{1}{N}\sum_{n}(T^n_{\mathrm{tem}}\circ I_n)$ is the implicit template by averaging warped input images \cite{vandemeulebroucke2011spatiotemporal}, $\lambda_0$ and $\lambda_1$ are the weights for smoothness and cyclic regularization, respectively. The cyclic regularization term will only be present if the relative motion in the image group is periodic or symmetric. The objective of the iterative optimization then becomes finding the optimal transformation $T^n_{\mathrm{tem}}$ that aligns every image in the group to a template image while keeping the deformation field smooth and cyclically consistent. 
The inverse transformation $T_n^{\mathrm{tem}}$ that maps the same anatomical locations in the input image to the implicit template is determined from a fixed-point method \cite{chen2008simple}. The transformation mapping between the $n$th and $m$th image $T_m^n$ can be calculated using the composition of the deformation field: $T_m^n(x)=T^n_{\mathrm{tem}}(T_m^{\mathrm{tem}}(x))$.

\begin{figure}[htb!]
    \centering
    \includegraphics[width=0.9\columnwidth]{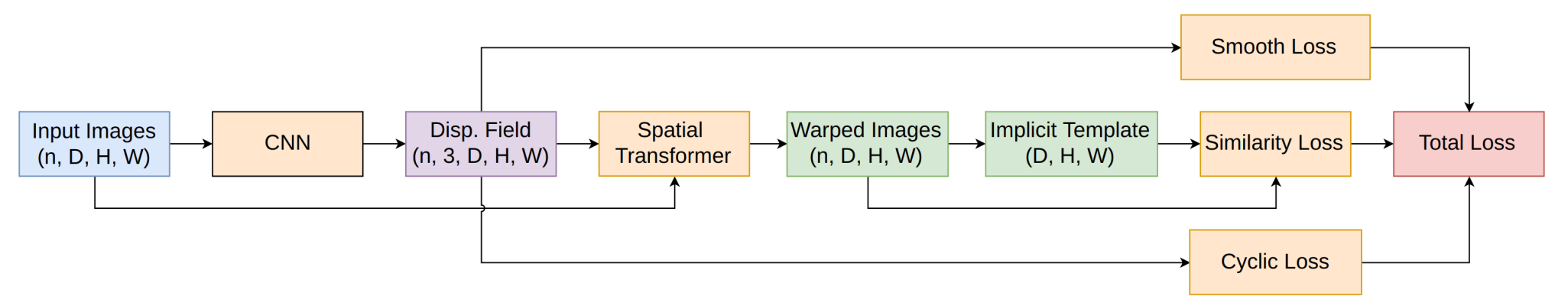}
    \caption{Flowchart of GroupRegNet. The expression (n, D, H, W) represents the number of images in the group and the spatial dimensions of the image. }
    \label{fig:flowchart}
\end{figure}

Figure \ref{fig:flowchart} illustrates the components and data flowing of GroupRegNet. 
As compared to the common structure of a learning-based method VoxelMorph \cite{balakrishnan2019voxelmorph}, GroupRegNet uses similar components including a CNN (to be explained in the later subsections), a spatial transformer (implemented as a 3D linear interpolation), a similarity loss, a cyclic loss, and a smoothness loss. The input images are processed by the CNN to directly estimate the displacement fields. Existing methods in the literature explicitly select the reference and moving images to form a pair and then warp the moving image to the reference image. By contrast, in GroupRegNet, the input images in the group are first stacked in the channel dimension before feeding into the neural network, and the computed transformation then aims to warp the input image into the common space of the template image. 
It should be noted that CNN's output is the displacement field $D^n_{\mathrm{tem}}(x)$ instead of the transformation field $T^n_{\mathrm{tem}}(x)$, which are related through $T^n_{\mathrm{tem}}(x)=D^n_{\mathrm{tem}}(x)+x$. 
The details of the components in this flowchart are further elaborated in the next subsections. 

\subsection{Network design}

\begin{figure}[htb!]
    \centering
    \includegraphics[width=0.8\columnwidth]{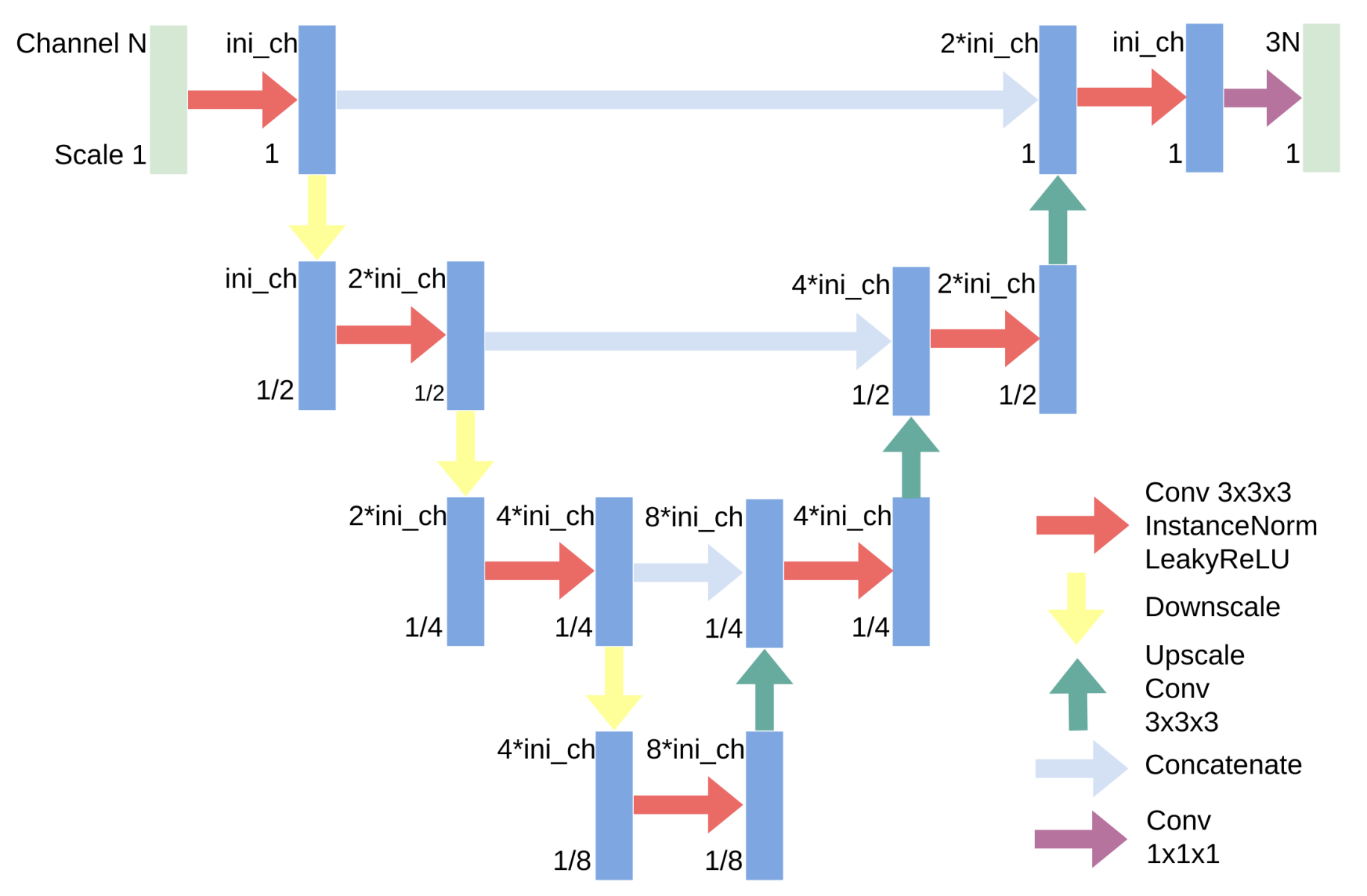}
    \caption{Detailed structure of the CNN sub-network. The overall design is similar to U-Net with modifications. The rectangle blocks represent the feature maps with denoted number of channels (top) and image scale (bottom). }
    \label{fig:cnn}
\end{figure}

The CNN model consists of convolution, downscale/upscale, and skip connection. The detailed structure of the CNN is shown in figure \ref{fig:cnn}. The overall structure is the same as U-net, which is used by most medical image registration networks. However, several changes have been made to meet the need of the one-shot groupwise registration. 
\begin{enumerate}
    \item In the original U-net, the downscale and upscale layers are implemented by max-pooling and transposed convolution. They are replaced by a more straightforward interpolation layer to convert the scales between feature maps. 
    \item The number of batches will always be one since only one group of images will be fed into the network during the optimization process. Therefore, the batch normalization is replaced by the instance normalization. 
    \item The two consecutive sets of convolution-normalization-activation operations are reduced to one. This change increases efficiency without impairing the performance. The leaky rectified activation layer is used instead of the original rectified linear activation(ReLU). 
    \item Due to the size limitation of the common video memory, the input image is downscaled to a lower resolution before being fed into the CNN. The output displacement field $D^n_{\mathrm{tem}}$ is then upscaled to the original resolution to warp the input images. The scale used in this work is $0.5$.
\end{enumerate}

\subsection{Loss functions}
The local normalized cross-correlation (NCC) coefficient is adopted to measure the similarity loss $L_{\mathrm{simi}}$ between the template and warped input images for its robustness against noise and intensity shift. Let $\bar{f}(x)=\sum_{x_i} f(x_i)/n^3$ and $\hat{f}(x)=\sum_{x_i} (f(x_i)-\bar{f}(x))^2$ denote the local mean and variance images, respectively, where $x_i$ loops over a cubic volume with a size $n^3$ around the voxel $x$, with $n=5$ in the current implementation. The NCC coefficient between the two images is calculated using
\begin{equation}\label{eq:ncc}
NCC(f,g)=\frac{1}{|\Omega|}\sum_{x\in\Omega}\frac{\sum_{x_{i}}(f(x_i)-\bar{f}(x))(g(x_i)-\bar{g}(x))}{\sqrt{\hat{f}(x)\hat{g}(x)}}.
\end{equation}
Accordingly, the similarity loss $L_{\mathrm{simi}}$ is the average negative NCC coefficient between an individual warped input image and the template image
\begin{equation}\label{eq:loss_simi}
L_{\mathrm{simi}}(T^N_{\mathrm{tem}}\circ I^N,I_{tem})=-\frac{1}{N}\sum_{n}NCC(T^n_{\mathrm{tem}}\circ I_n,I_{tem}).
\end{equation}
$L_{\mathrm{simi}}$ is in the range of $[-1,1]$ for which a lower value indicates a higher similarity. 

The smoothness regularization loss $L_{\mathrm{smo}}$ encourages a smooth and realistic transformation, which accounts the displacement field gradient and the gradient of the image \cite{liu2020learning}: 
\begin{equation}\label{eq:smooth}
L_{\mathrm{smo}}(D^N_{\mathrm{tem}},I_{tem})=\frac{1}{3N|\Omega|}\sum_{n,x\in\Omega,i\in X,Y,Z}(\Vert\nabla_iD^n_{\mathrm{tem}}(x)\Vert_1\exp(-|\nabla_iI_{tem}(x)|)).
\end{equation}
Here $\nabla_iD^n_{\mathrm{tem}}(x)$ is the partial derivative of the displacement field with respect to axis $i$, which is approximated by a forward difference. Our initial choice of the smoothness term was the isotropic total variation \cite{vishnevskiy2016isotropic}. The current term in equation \ref{eq:smooth} slightly increases the accuracy and efficiency by promoting the consistency between the gradient of displacement and the gradient of image intensity to preserve the edges.

An optional cyclic consistent regularization loss is used if  deformation fields in the group are periodic or symmetric, such as those present in a respiratory-binned 4D-CT. 
This loss reduces the sum of displacements from one location in the template to all corresponding locations in the input images so that the estimated template is at the center of all input images in the image manifold. 
An alternative cyclic loss is to reduce the composition of transformations through the cycle of motion. However, in practice, it is computationally expensive to implement. 
The smoothness and cyclic consistent loss are both positive values, and a lower loss represents the higher smoothness or consistency of the deformation field, respectively. 

\begin{equation}\label{eq:cyclic}
L_\mathrm{cyc}(T^N_{\mathrm{tem}})=\sqrt{\frac{1}{3|\Omega|}\sum_{x\in\Omega,i\in X,Y,Z}(\sum_nT^n_{\mathrm{tem},i}(x))^2}.
\end{equation}

\subsection{One-shot learning and convergence criterion}

The one-shot learning strategy is used in GroupRegNet to eliminate the requirement of abundant training data. The input images in the group are stacked in the channel dimension, then fed into the neural network to derive the current total loss and to update the weights iteratively through backpropagation. The weights in CNN are independently initialized at the beginning of each iterative registration process. In this sense, the one-shot strategy is similar to the iterative optimization in the variational registration. 

\begin{figure}[htb!]
    \centering
    \includegraphics[width=0.4\columnwidth]{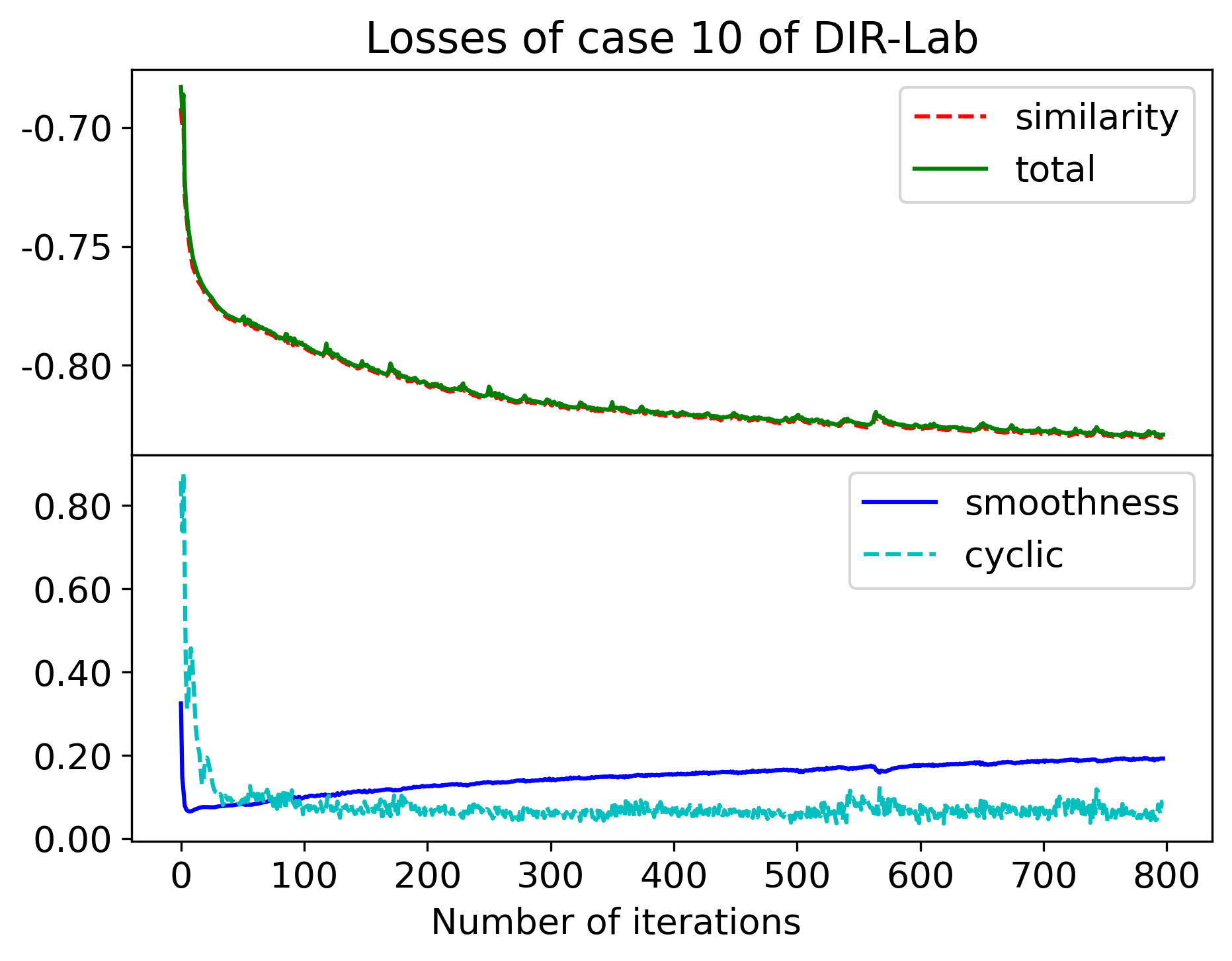}
    \caption{Example of a convergence curve: similarity $L_{\mathrm{simi}}$, smoothness $L_{\mathrm{smo}}$, cyclic $L_{\mathrm{cyc}}$, and total loss $L_{\mathrm{total}}$ vs. number of iteration for case 10 of the DIR-Lab dataset. 
    }
    \label{fig:simi_loss_iteration}
\end{figure}

After each iteration, a set of convergence criteria is evaluated to determine whether the iterative process should be terminated. The main criterion is the standard deviation of the recent similarity losses. A list of $N_{\mathrm{stop}}$ latest similarity losses is maintained. A lower standard deviation of this list indicates that a more stable solution has been reached. More specifically, the optimization will stop if 
\begin{enumerate}
    \item The standard deviation $\sigma$ of $N_{\mathrm{stop}}$ latest similarity losses is less than the threshold $\sigma_{\mathrm{stop}}$.
    \item Current similarity loss is not smaller than the previous minimum similarity loss and not larger than the previous minimum plus $\sigma_{\mathrm{stop}}/3$. 
    \item The number of computed iterations should be larger than a predefined value $N_{\mathrm{iter}}$. 
\end{enumerate}
The parameter $N_{\mathrm{stop}}$, $\sigma_{\mathrm{stop}}$, and $N_{\mathrm{iter}}$ are empirically determined to be $100$,  $0.0007$, and $200$ , and they are kept the same for all experiments. The determined displacement field $D^N_{\mathrm{tem}}$ is the output from the CNN of the last iteration. 
For all evaluated cases, this set of criteria and parameters have proved to be able to overcome the local minimum while avoiding prolonged computation. One example of the convergence curve of different losses vs. the number of iterations is shown in figure \ref{fig:simi_loss_iteration}. 

\subsection{Implementation details}
The proposed algorithm is implemented in PyTorch. 
The Adam optimizer with the learning rate of 0.01 is used for optimization. The number of downscales in CNN is set to 3 and the initial number of channels is 32. In this setting, the total number of trainable parameters in the network is 2.4 million. The default Kaiming initialization method is used for all convolutional layers. The regularization terms $\lambda_0$ and $\lambda_1$ are empirically set to $1\times10^{-3}$ and $1\times10^{-2}$, respectively. Computations are conducted on an 8-core CPU AMD Ryzen 3700X with a Nvidia 2080Ti GPU. To facilitate future research, the source code is available at \url{https://github.com/vincentme/GroupRegNet}.

\section{Experiments}
\subsection{Datasets}
To quantitatively evaluate the accuracy of GroupRegNet, the publicly available 4D-CT dataset DIR-Lab \cite{castillo2009framework} was used. This dataset provides 10 thorax 4D-CT scans, each consisting of 10 respiratory-binned phases. Three hundred pairs of corresponding landmarks in the lung were manually delineated by an expert at phases of end-inhalation (EI) and end-exhalation (EE). Two additional observers annotated part of the landmarks with the reported inter-observer variance ranged from $0.70\pm\SI{0.99}{mm}$ to $1.03\pm\SI{2.19}{mm}$. In addition, 75 sets of landmarks were delineated in all expiratory phase images, i.e. T00, T10, to T50. 

The registration accuracy was evaluated by comparing the Euclidean distance, i.e., target registration error (TRE), between the deformed landmarks using the determined deformation fields and annotated landmarks. Note that the 300 pairs of landmarks provided by DIR-Lab suffer from two limitations. First, the number and density of landmarks are limited. Second, the accuracy of landmarks is only at the voxel level. \citet{fu2019automatic} recently proposed an automatic method that can generate a large amount of matching landmarks (1886 pairs on average) evenly distributed in the lung region with subvoxel-level accuracy (average TRE of $0.47\pm\SI{0.45}{mm}$). 
Therefore, these dense matching landmarks were also used in this study. The landmarks provided by DIR-Lab and by \citet{fu2019automatic} are denoted by Landmark300 and LandmarkDense, respectively. 


Another dataset, the point-validated pixel-based breathing thorax (POPI) from \cite{vandemeulebroucke2011spatiotemporal} was also used to quantitatively evaluate the registration algorithm. This dataset consists of six respiratory phase-binned 4D-CT. About 100 pairs of corresponding landmarks per case at EI and EE phases were created by a semi-automatic approach. 

\begin{figure}[htb!]
    \centering
    \includegraphics[width=0.7\columnwidth]{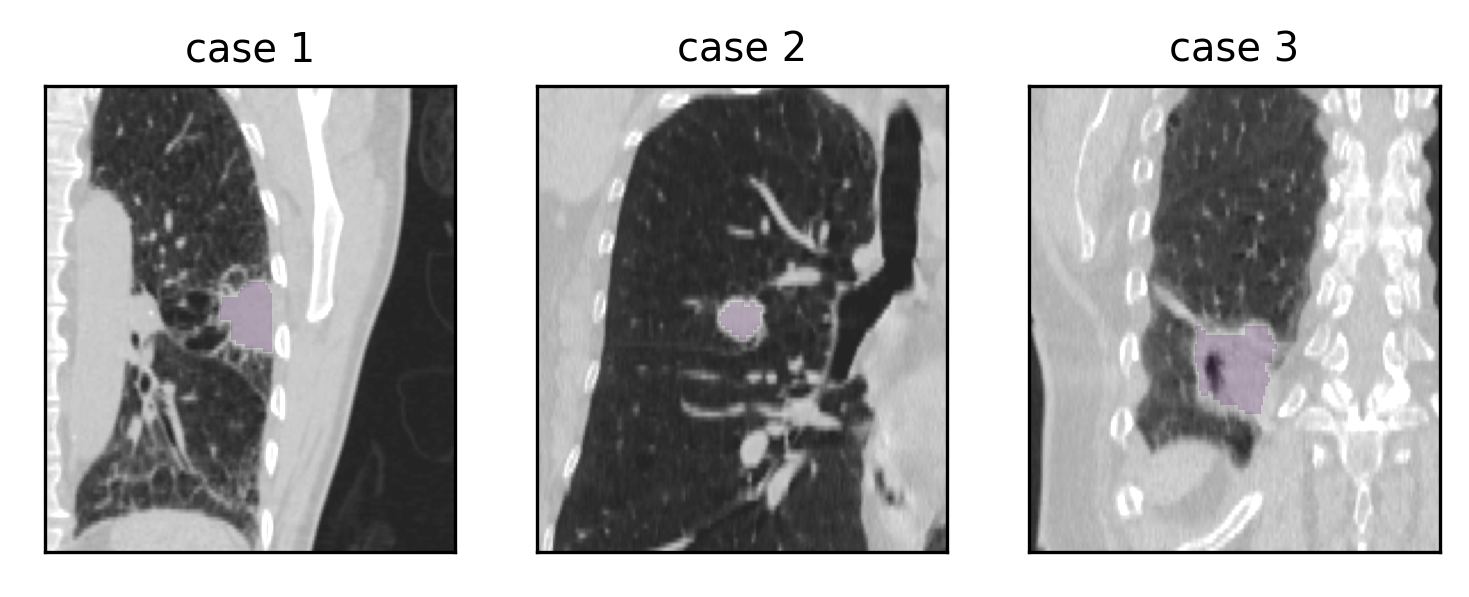}
    \caption{Sizes, shapes, and locations of the contoured tumor targets, shown in violet shade, in coronal views of the EI phases of three patient cases. }
    \label{fig:target_location}
\end{figure}

In addition, 4D-CT scans of three lung cancer patients were obtained in the authors' department to quantitatively evaluate tumor tracking using GroupRegNet. The 4D-CT dataset for each patient consists of 10 3D CT volumes representing ten respiratory phases (T00, T10, ..., T90), from the end-inhalation phase (T00) to the end-exhalation phase (T50) then back (T90). The image voxel size was $1.18\times1.18\times\SI{2}{mm^3}$. The tumors were manually contoured by a trained medical physicist on every respiratory phase for each patient. The sizes, shapes, and locations of the tumor targets varied among patients as shown in figure \ref{fig:target_location}. For case 2, the tumor was in the center of the right lung. For cases 1 and 3, the tumors were next to the chest wall and the spine, respectively. To track the tumor motion, the 4D-CT images were first registered using GroupRegNet, then the segmentation mask of the EI phase was warped to other phases with the computed DVFs. The GroupRegNet tumor tracking accuracy was evaluated by computing the average and standard deviation of the Dice coefficients, the distances between the centers of mass, and the 95\% Hausdorff distances using the manually contoured tumor target masks as reference. 

\subsection{Prepossessing}
To reduce computation time and improve convergence, the input images were cropped to the bounding box that encompassed the landmarks in all phases plus an 8-voxel margin in all directions. In the cases of tumor target tracking, a 50-voxel margin was added to all directions of the boundingbox of the tumor target in the EI phase. The CT image intensity was approximately normalized to the range of [-1,1] after dividing by 1000. The input images were not spatially resampled, segmented, or vessel enhanced before being fed into GroupRegNet. 

\section{Results}
\subsection{Accuracy evaluated on landmarks}

\begin{table*}[htb!]
\centering
\caption{Comparison of TREs (mean$\pm$std in \SI{}{mm}): GroupRegNet vs. other learning-based and conventional DIR methods using the DIR-Lab dataset evaluated by (a) Landmark300 and (b) LandmarkDense.}
\label{tab:tre_dirlab}
    \centering
    \begin{subfigure}[c]{\textwidth}
        \centering
\resizebox{\textwidth}{!}{
\begin{tabular}{r|ccccccccc}
\toprule
case    & before reg.    & GroupRegNet      & LungRegNet\cite{fu2020lungregnet} & Fechter\cite{fechter2020one} & MJ-CNN\cite{jiang2020multi} & GDL-FIRE\cite{sentker2018gdl} & Fu\cite{fu2018adaptive} & Bartlomiej\cite{papiez2018fast} & pTVreg\cite{vishnevskiy2016isotropic}\\
\midrule
1        & $3.89\pm2.78$  & $1.02\pm0.51$ & $0.98\pm0.54$                     & $1.21\pm0.88$                & $1.20\pm0.63$               & $1.20\pm0.60$                 & $1.06\pm0.50$           & $0.90\pm1.0$  & $0.80\pm0.89$                \\
2        & $4.34\pm3.90$  & $1.04\pm0.49$ & $0.98\pm0.52$                     & $1.13\pm0.65$                & $1.13\pm0.56$               & $1.19\pm0.63$                 & $1.09\pm0.57$           & $0.94\pm1.0$      & $0.77\pm0.90$            \\
3        & $6.94\pm4.05$  & $1.24\pm0.71$ & $1.14\pm0.64$                     & $1.32\pm0.82$                & $1.30\pm0.70$               & $1.67\pm0.90$                 & $1.51\pm1.00$           & $1.06\pm1.1$      & $0.92\pm1.07$            \\
4        & $9.83\pm4.86$  & $1.43\pm0.97$ & $1.39\pm0.99$                     & $1.84\pm1.76$                & $1.55\pm0.96$               & $2.53\pm2.01$                 & $1.73\pm1.55$           & $2.53\pm3.2$     & $1.30\pm1.27$          \\
5        & $7.48\pm5.51$  & $1.41\pm1.22$ & $1.43\pm1.31$                     & $1.80\pm1.60$                & $1.72\pm1.28$               & $2.06\pm1.56$                 & $1.80\pm1.63$           & $1.31\pm1.5$     & $1.13\pm1.42$          \\
6        & $10.89\pm6.96$ & $1.31\pm0.72$ & $2.26\pm2.93$                     & $2.30\pm3.78$                & $2.02\pm1.70$               & $2.90\pm1.70$                 & $2.25\pm2.61$           & $1.89\pm1.9$      & $0.78\pm0.92$             \\
7        & $11.02\pm7.42$ & $1.28\pm0.65$ & $1.42\pm1.16$                     & $1.91\pm1.65$                & $1.70\pm1.03$               & $3.60\pm2.99$                 & $1.41\pm0.98$           & $1.52\pm1.4$    & $0.79\pm0.91$               \\
8        & $14.99\pm9.00$ & $1.33\pm1.08$ & $3.13\pm3.77$                     & $3.47\pm5.00$                & $2.64\pm2.78$               & $5.29\pm5.52$                 & $3.53\pm5.70$           & $1.87\pm2.3$      & $1.00\pm1.29$             \\
9        & $7.92\pm3.97$  & $1.30\pm0.69$ & $1.27\pm0.94$                     & $1.47\pm0.85$                & $1.51\pm0.94$               & $2.38\pm1.46$                 & $2.31\pm1.88$           & $1.37\pm1.1$       & $0.91\pm0.95$            \\
10       & $7.30\pm6.34$  & $1.22\pm0.63$ & $1.93\pm3.06$                     & $1.79\pm2.24$                & $1.79\pm1.61$               & $2.13\pm1.88$                 & $1.18\pm1.97$           & $1.27\pm1.4$      & $0.82\pm0.97$             \\
ave.     & $8.46\pm5.48$  & $1.26\pm0.77$ & $1.59\pm1.58$                     & $1.83\pm2.35$                & $1.66\pm1.44$               & $2.50\pm1.16$                 & $1.78\pm1.83$           & $1.47\pm1.6$   & $0.92\pm1.06$               \\
ave. RMSE   & $10.08$  & $1.48$ & $2.24$                     & $2.98$                & $2.20$               & $2.76$                 & $2.55$           & $2.2$     & $1.41$             \\
\bottomrule
\end{tabular}
}
        \caption{Landmark300}
    \end{subfigure}\\
    \begin{subfigure}[c]{\textwidth}
        \centering
\resizebox{0.45\textwidth}{!}{
\begin{tabular}{r|ccc}
\toprule
case    & before reg.     & GroupRegNet    & pTVreg\cite{vishnevskiy2016isotropic}   \\
\midrule
1       & $3.43\pm2.86$   & $0.59\pm0.33$  & $0.32\pm0.17$ \\
2       & $4.67\pm4.23$   & $0.56\pm0.36$  & $0.38\pm0.22$ \\
3       & $5.55\pm4.08$   & $0.71\pm0.37$ & $0.41\pm0.23$  \\
4       & $7.55\pm5.11$   & $0.70\pm0.35$  & $0.56\pm0.55$ \\
5       & $4.91\pm4.84$   & $0.65\pm0.36$  & $0.47\pm0.30$ \\
6       & $9.30\pm7.46$   & $0.96\pm0.57$  & $0.70\pm1.63$ \\
7       & $8.18\pm6.73$   & $0.78\pm0.40$ & $0.48\pm0.25$  \\
8       & $8.58\pm6.71$   & $0.81\pm0.43$  & $0.61\pm1.68$ \\
9       & $5.81\pm3.77$   & $0.83\pm0.45$   & $0.48\pm0.25$\\
10      & $6.12\pm5.31$   & $0.77\pm0.43$  & $0.44\pm0.25$ \\
ave. & $6.41\pm5.11$   & $0.74\pm0.41$ & $0.49\pm0.55$ \\
ave. RMSE & $8.20$   & $0.85$  &  $0.74$  \\
\bottomrule
\end{tabular}
}
        \caption{LandmarkDense}
    \end{subfigure}

\end{table*}

The accuracy of GroupRegNet was compared with seven recently published methods on the DIR-Lab dataset, as shown in table \ref{tab:tre_dirlab}. The landmarks in EI phase (phase T00) were deformed to EE phase (phase T50) according to the calculated DVFs, and then compared to the annotated landmarks in EE phase to derive the TREs. GroupRegNet and pTVreg were evaluated on both LandmarkDense and Landmark300, while other methods only reported results on Landmark300. 

The average TRE of GroupRegNet was $1.26\pm\SI{0.77}{mm}$,  evaluated on Landmark300, which was lower than most of the surveyed methods, and comparable to pTVreg \cite{vishnevskiy2016isotropic}, which is the top method listed on the DIR-Lab website. The average root mean square error (RMSE) of GroupRegNet and pTVreg were at least $30\%$ smaller than other methods. GroupRegNet performed particularly better for cases with large deformations (e.g., cases 6, 7 and 8). It should also be noted that the variance of the TREs using GroupRegNet was even less or at least equal to the inter-observer variance, suggesting that its accuracy was superior to that of manual annotations in most regions. 

When evaluated using LandmarkDense, the average TRE and RMSE of GroupRegNet were $0.74\pm\SI{0.41}{mm}$ and $\SI{0.85}{mm}$, respectively, demonstrating a sub-millimeter accuracy. 
The average RMSEs were similar comparing GroupRegNet vs. pTVreg while the former usually yielded smaller standard deviations but slightly larger average TREs. Note that the standard deviations of pTVreg in cases 6 and 8 were unexpectedly large, which was not observed in GroupRegNet.

\begin{figure}[htb!]
    \centering
    \begin{subfigure}[c]{0.4\textwidth}
        \centering
        \includegraphics[width=\textwidth]{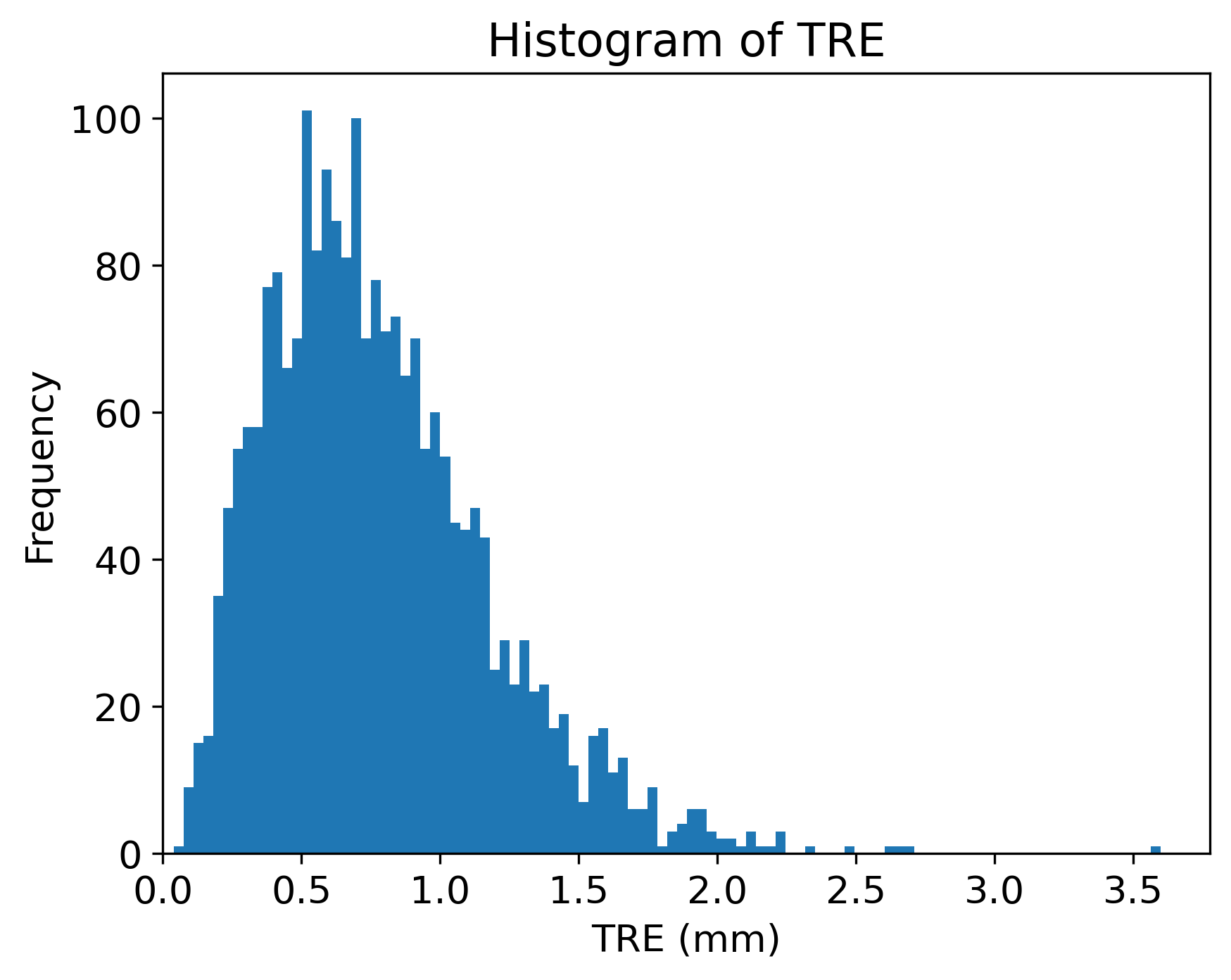}
        \caption{}
    \end{subfigure}~
    \begin{subfigure}[c]{0.35\textwidth}
        \centering
        \includegraphics[width=\textwidth]{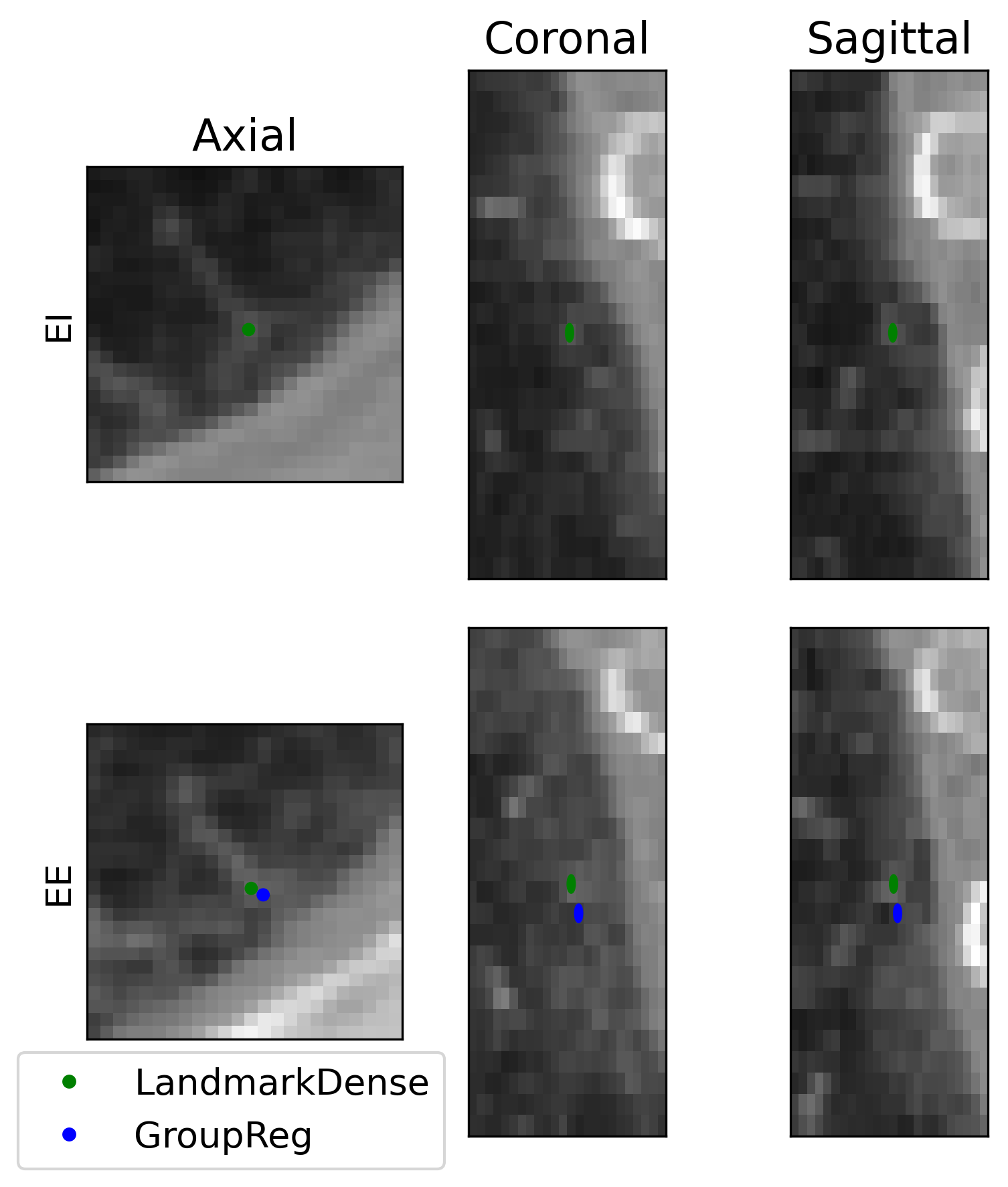}
        \caption{}
    \end{subfigure}
    \caption{Accuracy of GroupRegNet evaluated on LandmarkDense in case 7 of DIR-Lab. (a) histogram of TREs, (b) the location of the worst point determined by GroupRegNet in phases EI and EE. } 
    \label{fig:tre_histogram}
\end{figure}

The TRE histogram for GroupRegNet in case 7 is shown in figure \ref{fig:tre_histogram}(a) where the percentage of the TREs below $\SI{1}{mm}$, $\SI{1.5}{mm}$, and $\SI{2}{mm}$ are $75\%$, $94\%$, and $99\%$, respectively. The worst point with a TRE of $\SI{3.6}{mm}$ is shown in figure \ref{fig:tre_histogram}(b). The relatively large error was likely caused by the low signal-to-noise ratio and the rapidly changing displacement in this region. 
Figure \ref{fig:compare_reg_disp} provides a typical example of the DIR results. Most structures align well as shown in the red/cyan superimposed image post registration.

\textbf{\begin{figure}[htb!]
    \centering
    \includegraphics[width=0.8\columnwidth]{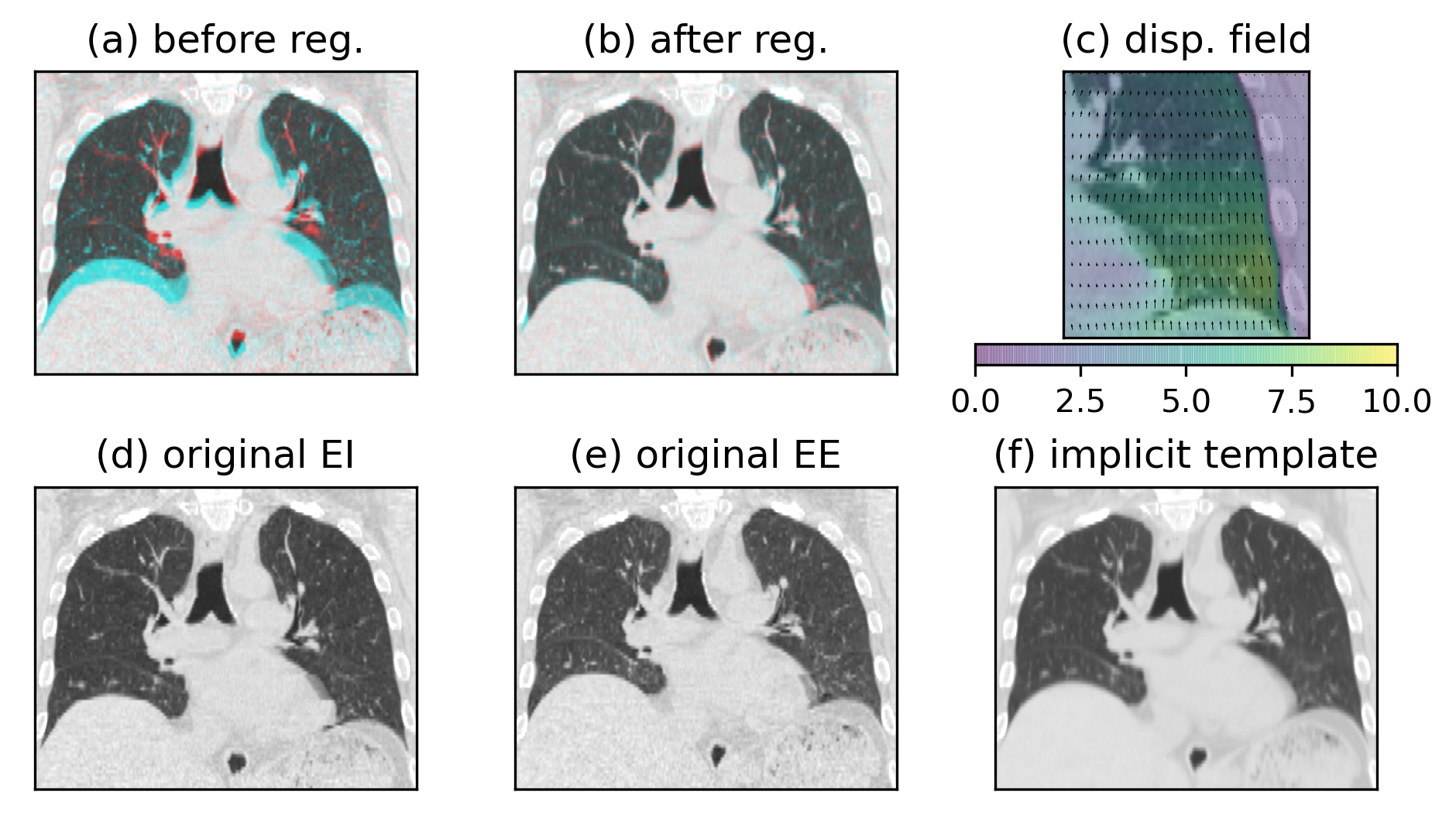}
    \caption{Example of the registration result of a coronal slice of DIR-Lab case 7. The red/cyan superimposed images of phases EI and EE (a) before and (b) after registration. (c) The colormap and vector plot show the magnitude and direction of a portion of the determined displacement field from phase EI to EE. (d), (e), and (f) are the images of phases EI, EE, and determined implicit template, respectively.}
    \label{fig:compare_reg_disp}
\end{figure}}

\begin{table*}[htb!]
\centering
\caption{Comparison of TREs (mean$\pm$std in \SI{}{mm}) between GroupRegNet and other learning-based or conventional methods on POPI dataset.}
\label{tab:tre_popi}
\resizebox{0.8\textwidth}{!}{
\begin{tabular}{@{}r|ccccc@{}}
\toprule
case    & dimensions  & before reg.    & GroupRegNet      & Fechter\cite{fechter2020one} & GDL-FIRE\cite{sentker2018gdl} \\ \midrule
1       & 512x512x141 & $5.90\pm2.73$  & $1.10\pm0.59$ & $1.09\pm0.68$                & $1.34\pm0.74$                 \\
2       & 512x512x169 & $14.04\pm7.20$ & $1.27\pm0.93$ & $2.71\pm3.28$                & $2.98\pm2.38$                 \\
3       & 512x512x170 & $7.67\pm5.05$  & $0.92\pm0.51$ & $1.40\pm1.54$                & $1.57\pm1.01$                 \\
4       & 512x512x187 & $7.33\pm4.89$  & $0.88\pm0.47$ & $1.17\pm1.83$                & $1.64\pm1.62$                 \\
5       & 512x512x139 & $7.09\pm5.08$  & $1.01\pm0.81$ & $1.30\pm0.97$                & $1.62\pm1.09$                 \\
6       & 512x512x161 & $6.68\pm3.68$  & $0.97\pm0.51$ & $1.27\pm0.95$                & $1.26\pm0.73$                 \\
ave. &             & $8.12\pm4.77$  & $1.03\pm0.64$ & $1.49\pm1.54$                & $1.74\pm1.26$                 \\
ave. RMSE &             & $9.42$  & $1.21$ & $2.14$                & $2.15$                 \\
\bottomrule
\end{tabular}
}
\end{table*}

The results evaluated on POPI are shown in table \ref{tab:tre_popi}. GroupRegNet reduced the original TRE from $8.12\pm\SI{4.77}{mm}$ to $1.03\pm\SI{0.64}{mm}$. Comparing to the results from \citet{fechter2020one} and GDL-FIRE \cite{sentker2018gdl}, the average RMSE was reduced by $44\%$. 

\begin{table*}[htb!]
\centering
\caption{Comparison of TREs (mean$\pm$std in \SI{}{mm}) of GroupRegNet on different target phase images from the DIR-Lab dataset using 75 landmarks. The result of pTVreg on phase T50 is included for reference. }
\label{tab:tre_dirlab75}
\resizebox{0.8\textwidth}{!}{
\begin{tabular}{r|cccccc}
\toprule
case & T10           & T20           & T30           & T40           & T50           & pTVreg T50           \\
\midrule
1    & $0.38\pm0.29$ & $0.95\pm0.65$ & $1.27\pm0.60$ & $1.22\pm0.64$ & $1.15\pm0.57$  & $0.92\pm0.49$ \\
2    & $0.97\pm0.77$ & $0.94\pm0.58$ & $0.93\pm0.56$ & $0.94\pm0.52$ & $1.00\pm0.53$ & $0.92\pm0.49$ \\
3    & $1.17\pm0.79$ & $1.12\pm0.58$ & $1.10\pm0.59$ & $1.23\pm0.61$ & $1.21\pm0.62$ & $1.01\pm0.50$  \\
4    & $1.10\pm0.60$ & $1.32\pm0.80$ & $1.41\pm0.83$ & $1.55\pm1.24$ & $1.39\pm0.98$ & $1.28\pm0.91$ \\
5    & $1.47\pm1.08$ & $1.20\pm0.57$ & $1.23\pm0.91$ & $1.21\pm0.62$ & $1.57\pm1.86$ & $1.34\pm1.78$ \\
6    & $1.07\pm0.89$ & $1.79\pm1.78$ & $1.64\pm1.70$ & $1.54\pm1.36$ & $1.31\pm0.83$ & $1.04\pm0.76$ \\
7    & $0.99\pm0.79$ & $1.42\pm1.16$ & $1.58\pm1.07$ & $1.27\pm0.81$ & $1.35\pm0.65$ & $0.94\pm0.49$ \\
8    & $1.04\pm0.52$ & $1.42\pm1.22$ & $1.32\pm1.09$ & $1.73\pm2.13$ & $1.50\pm1.78$ & $1.22\pm1.74$ \\
9    & $1.14\pm0.62$ & $1.17\pm0.70$ & $1.22\pm0.67$ & $1.37\pm0.70$ & $1.31\pm0.77$ & $1.09\pm0.78$ \\
10   & $1.16\pm0.93$ & $1.38\pm1.18$ & $1.64\pm1.36$ & $1.23\pm0.63$ & $1.16\pm0.56$ & $0.91\pm0.43$ \\
\bottomrule
\end{tabular}
}
\end{table*}

\begin{figure}[htb!]
    \centering
    \includegraphics[width=0.8\columnwidth]{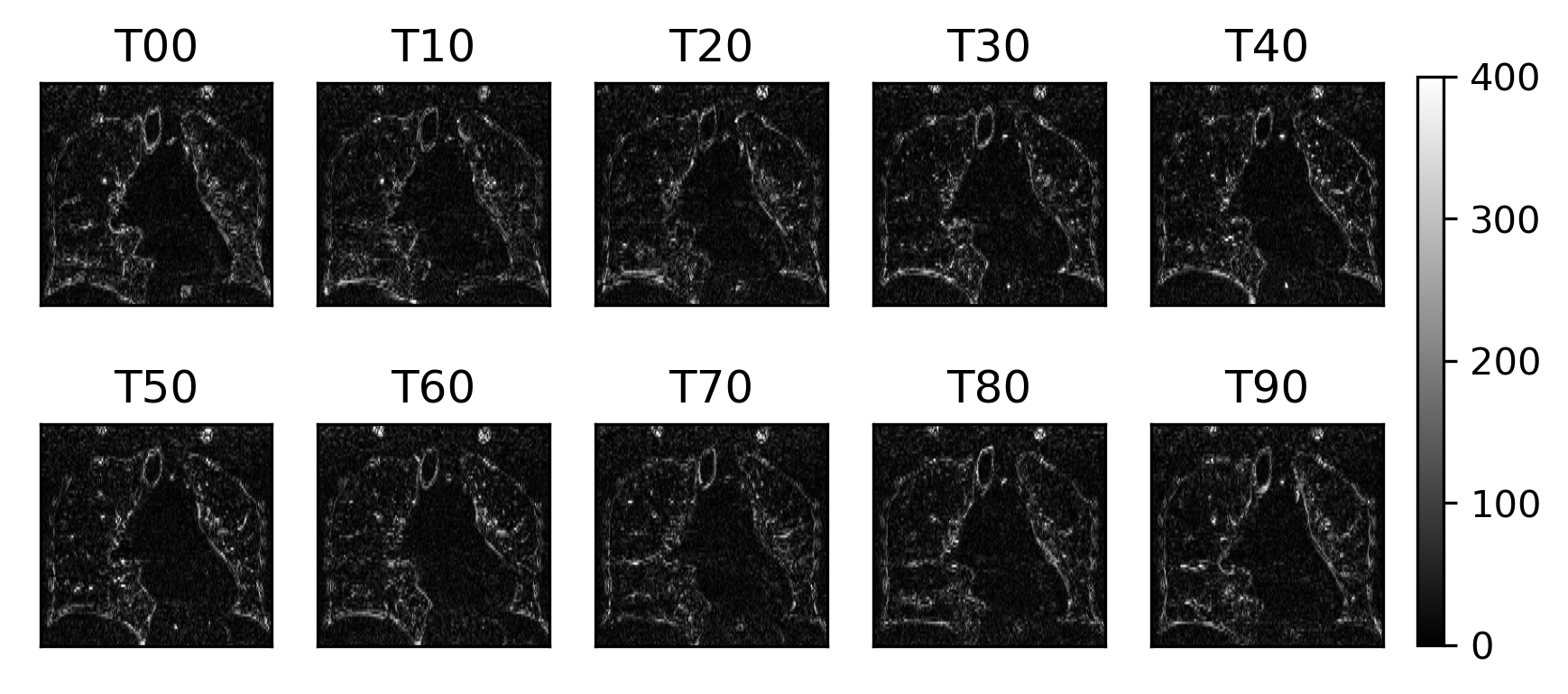}
    \caption{Intensity-difference map between each phase and warped template in coronal view of DIR-Lab case 10. }
    \label{fig:diff}
\end{figure}

All previous evaluations were carried out between phases EI and EE. The 75 landmarks annotated on the expiratory phases of the DIR-Lab dataset were utilized to test whether there are large variations among different phases. The landmarks in phase T00 were deformed to other phases and then compared to manual annotations, as shown in table \ref{tab:tre_dirlab75}. The TREs of phases T10 and T50 were usually smaller than those of other phases, which could be attributed to the former having smaller deformations and the latter being more stable than the intermediate phases. In addition, the intensity difference maps between each phase and the warped template image via the reverse DVF $T_n^{\mathrm{tem}}$ are shown in figure \ref{fig:diff}. There was not a single intensity-difference map that was obviously better or worse than its counterpart, suggesting similar GroupRegNet performance regardless of phases.

The percentage of the negative determinant of the Jacobian matrix was calculated to evaluate the regularity of the deformation fields. For most DIR-Lab cases, the percentage was zero, except for case 8, which was $0.03\%$. For the POPI dataset cases, the percentage was $0.06\pm0.13\%$. These results indicate the determined deformation field was well regularized with a minimal percentage of the negative determinant of the Jacobian matrix. 

\subsection{Accuracy evaluated on segmentation contours}
\begin{figure}[htb!]
    \centering
    \includegraphics[width=0.8\columnwidth]{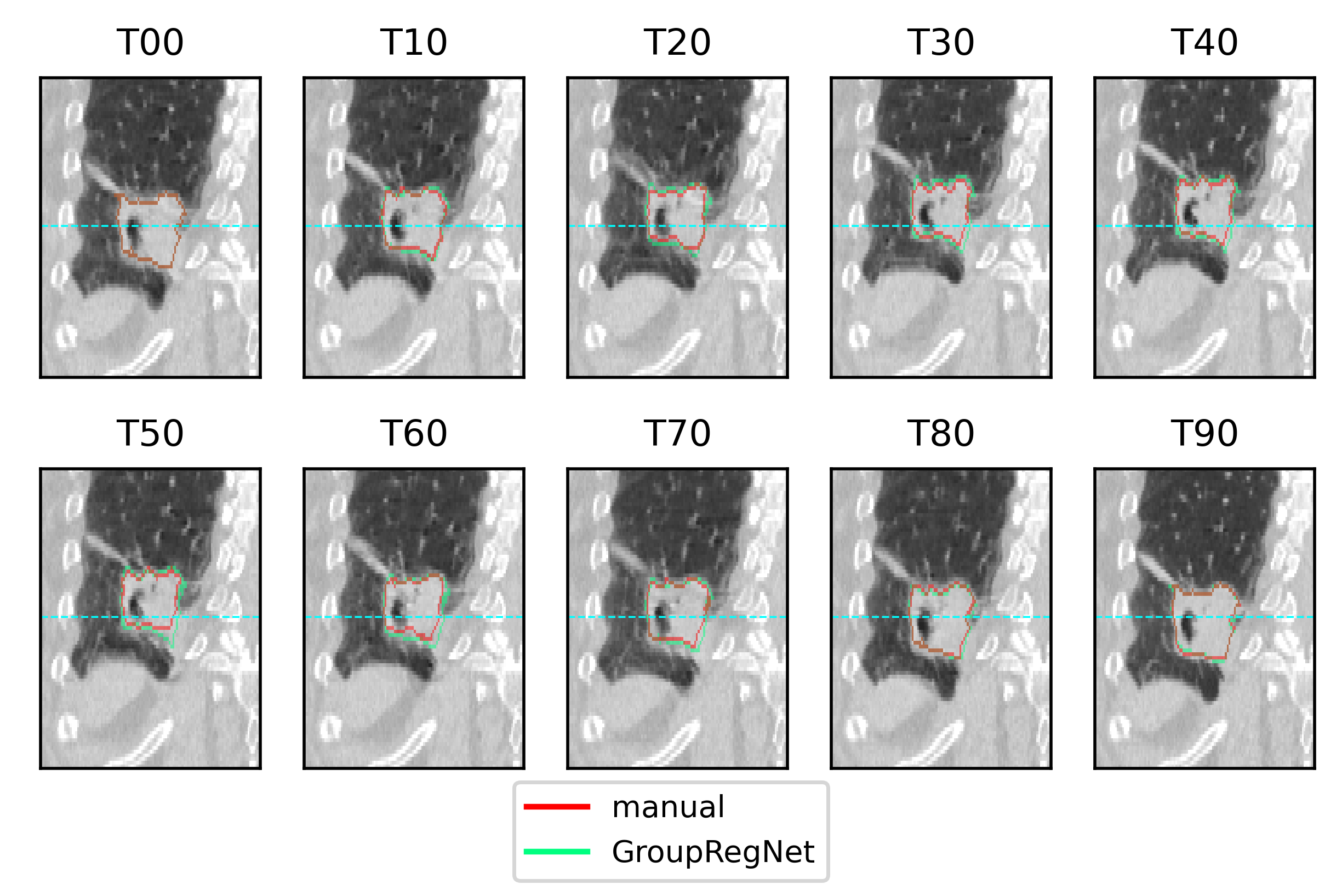}
    \caption{Comparison of the tracked targets in ten phases by GroupRegNet and manual contouring of case 3. The images are shown in coronal views, and the horizontal line in each figure is at the same height for visual reference. }
    \label{fig:tracked_target}
\end{figure}

As shown in table \ref{tab:segmentation}, the average and standard deviation of the Dice coefficients, distances between the centers of mass, and 95\% Hausdorff distances of the warped masks were computed with the manual contours as the reference. After registration, the average Dice coefficient was increased from $0.8$ to $0.9$, the distance between the centers of mass was reduced by 81\% to $\sim\SI{1}{mm}$, and the 95\% Hausdorff distance was reduced by 50\% to $<\SI{3}{mm}$. Figure \ref{fig:tracked_target} provides a visual example of the tracked target in different phases of case 3. 

\begin{table*}[htb!]
\centering
\caption{Comparison of the Dice coefficients, distances between the centers of mass, and 95\% Hausdorff distances of segmentation contours of the tumor targets before and after registration. }
\label{tab:segmentation}
\resizebox{0.7\textwidth}{!}{\begin{tabular}{@{}r|cc|cc|cc@{}}
\toprule
case                 & \multicolumn{2}{c}{Dice coefficient}          & \multicolumn{2}{c}{distance between the centers of mass}  & \multicolumn{2}{c}{95\% Hausdorff distance} \\ 
\multicolumn{1}{l}{} & \multicolumn{2}{c}{mean$\pm$std}              & \multicolumn{4}{c}{mean$\pm$std (\SI{}{mm})} \\ 
                     & before          & after           & before          & after          & before          & after          \\ \midrule
1                    & $0.789\pm0.070$ & $0.903\pm0.006$ & $4.58\pm2.14$   & $1.10\pm0.32$  & $4.73\pm1.48$   & $2.32\pm0.01$  \\
2                    & $0.807\pm0.080$ & $0.913\pm0.015$ & $1.82\pm0.81$   & $0.37\pm0.16$  & $2.17\pm0.63$   & $1.33\pm0.38$  \\
3                    & $0.781\pm0.130$ & $0.887\pm0.043$ & $5.74\pm3.60$   & $0.78\pm0.31$  & $6.08\pm3.22$   & $2.84\pm0.87$  \\ \bottomrule
\end{tabular}
}
\end{table*}

\subsection{Computation variance and speed}

\begin{table*}[htb!]
\centering
\caption{Comparison of variance of repeatability error, TRE, and  computation speed from five repeated runs by GroupRegNet on selected cases. The input images were cropped, so the dimensions were smaller than the originals.}
\label{tab:variance_speed}
\resizebox{\textwidth}{!}{
\begin{tabular}{r|ccc|cccc}
\toprule
               & repeatability error & mean of TREs & std of TREs & cropped dimensions & num. of iter. & computation time      & time per iter.  \\
    & \multicolumn{3}{c}{mean$\pm$std (\SI{}{mm})} & & & \multicolumn{2}{c}{\SI{}{s}} \\
\midrule
DIR-Lab case 1  & $0.21\pm0.12$  & $0.59\pm0.02$ & $0.34\pm0.03$ & $240\times157\times83$   & $317\pm24$    & $265\pm18$   & $0.8$ \\
DIR-Lab case 6  & $0.41\pm0.21$ & $0.94\pm0.03$ & $0.55\pm0.03$ & $294\times184\times97$   & $764\pm55$    & $973\pm255$  & $1.3$ \\
POPI case 2   & $0.47\pm0.23$   & $1.29\pm0.02$ & $0.95\pm0.02$ & $271\times196\times116$  & $1073\pm151$  & $1792\pm250$ & $1.7$ \\
POPI case 5   & $0.37\pm0.17$ & $1.02\pm0.01$ & $0.84\pm0.02$ & $169\times128\times99$   & $712\pm78$    & $412\pm44$   & $0.6$ \\
\bottomrule
\end{tabular}
}
\end{table*}

Due to the stochastic nature of weights initialization in the neural network, concerns may arise with regard to optimization convergence and variance among multiple runs. 
In addition, the computation speed is important in practical applications. Two cases with relatively small and large motions from both datasets were repeatedly registered five times using GroupRegNet. The variance of the registration accuracy, number of iteration, and computation time are summarized in table \ref{tab:variance_speed}. The variance of the registration accuracy was evaluated in terms of repeatability errors and statistics of TREs. The former was calculated as the distance between the displaced landmarks and their average locations over five runs. Then the average and standard deviation of the repeatability error were computed over all landmarks and runs. The determined repeatability errors ranged from $\SI{0.2}{mm}$ to $\SI{0.5}{mm}$. Although the variances were not minimum, the standard deviation of the statistics of the TREs was at the level of $\SI{0.03}{mm}$, indicating similar accuracies of repeated runs. Furthermore, all registrations were completed without convergence issues. 

Computation time per iteration ranged from $\SI{0.6}{s}$ to $\SI{1.7}{s}$, and varied with image size and motion magnitude. The overall computation time was in the range of  few minutes to 30 minutes, which is not slow considering that all 10 phases were registered and all pairwise DVFs determined. 

\section{Discussion}
A new DIR method GroupRegNet is presented to register 4D medical images and to determine all pairs of dense DVFs. The results on two respiratory 4D-CT datasets suggest  that it is able to achieve state-of-the-art performance. This study is unique in that it has successfully combined and implemented implicit template groupwise registration and one-shot unsupervised learning approach. Although many components have been introduced in the literature, in this work they are strategically integrated, and the method outperforms many other complex and dedicated methods. For instance, figure \ref{fig:compare_reg_disp}(c) shows the DVF transition around the chest wall where the sliding motion was successfully revealed without additional dedicated steps such as segmentation or DVF decomposition\cite{fu2018adaptive}. The implicit template shown in figure \ref{fig:compare_reg_disp}(f) was successfully revealed by averaging the warped input images, which showed less noise compared to the original images. This is also an advantage of the implicit template groupwise registration method over the pairwise registration method; for the latter both the reference and moving images are inevitably corrupted by noise. 

From a broader perspective, GroupRegNet can be viewed as a mixture of conventional and learning-based methods. It follows the same iterative optimization process of the conventional approach and only uses the images to be registered as input. Furthermore, segmentation images, annotated landmarks, or deformation fields do not need to be provided to the neural network. GroupRegNet utilizes CNN as the motion model whose weights are learned through optimization. The performance improvement over the conventional approach can be attributed to the more expressive power of the deep neural network and to fewer assumptions in the DVF. Comparing to a typical training and inference procedure of learning-based methods, the one-shot learning strategy eliminates the requirement of abundant training images and annotations, thus improves the accuracy. The problem of over-fitting for the one-shot training strategy was not presented due to the well regularized total loss. 

GroupRegNet utilizes several concepts from previously published registration methods that include group-wise implicit template registration and one-shot unsupervised learning. \citet{fechter2020one} proposed the one-shot unsupervised learning approach but the deformation field computation between timely adjacent 3D images required segmented inputs and a complex coarse-to-fine patch processing.  \citet{wu2013estimating} introduced the implicit template into the paradigm of the classical registration approach that consists of explicitly defined key points and the Gaussian mixture model for motion modeling. By contrast, GroupRegNet organically integrates these components, and features a simpler network design, a minimal prepossessing step, and a straightforward registration process compared to other methods. 
Comparing to other learning-based methods, \citet{fu2020lungregnet} designed separate coarse and fine nets for large and small deformations, and also included a discriminator and a corresponding adversarial loss to regularize DVF. 
Moreover, both studies \cite{fu2020lungregnet,fechter2020one} ran into limited GPU memory issues so that they had to be trained using a patch-based approach, which was time-consuming and hard to learn the global relationship. \citet{fechter2020one} paid special attention to the smoothness regularization of the boundary voxels while \citet{fu2020lungregnet} excluded these voxels in the loss. GroupRegNet tackles this limitation differently where the input images to CNN were downscaled to reduce the size of the feature maps so that the model can run on a typical GPU. Furthermore, the output DVFs are upscaled to the original resolution to warp the input images and then to compute the similarity loss. 
The global representations are learned instead of using local features in the patch. The output DVFs and similarity regularization computation are computed at the original resolution. This approach is also better than conducting all  computation on a lower resolution, which loses the fine details of input images and reduces the accuracy of the DIR. 

\section{Conclusion}
In this paper, a groupwise one-shot learning neural network for 4D image registration was presented. The implicit template strategy was first integrated with the learning-based approach. The utilization of one-shot learning strategy eliminated the need for abundant training data. The simple network structure made the registration at the original resolution without breaking up the input images into patches. The accuracy of GroupRegNet in terms of average RMSE was better than that of the latest learning-based methods and comparable to the top conventional method. The performance of GroupRegNet is expected to be further improved with the addition of more complex networks and strategies, such as generative adversarial network and attention mechanism.

\section*{Acknowledgments}
This research was partially supported by the Agency for Healthcare Research and Quality (AHRQ) grant number R01-HS022888, National Institute of Biomedical Imaging and Bioengineering (NIBIB) grant R03-EB028427 and National Heart, Lung, and Blood Institute (NHLBI) grant R01-HL148210. 


\newcommand{\newblock}{}
\bibliographystyle{plainnat}
\bibliography{reference}

\end{document}